\documentclass[aps,pre,superscriptaddress,preprint]{revtex4-1}
\usepackage{eurosym}
\usepackage{amsfonts}
\usepackage{amsmath}
\usepackage{amssymb}
\usepackage{graphicx}
\usepackage{color}

\begin{document}

\title{Density Response from Kinetic Theory and Time Dependent Density
Functional Theory for Matter Under Extreme Conditions}
\author{James Dufty}
\affiliation{Department of Physics, University of Florida, P.O. Box 118435,
	Gainesville, Florida 32611-8435, USA}

\author{Kai Luo}
\author{S.B. Trickey}

\affiliation{Quantum Theory Project, Department of Physics and Department of Chemistry,\\
	University of Florida, P.O. Box 118435, Gainesville, Florida 32611-8435, USA}

\begin{abstract}
The density linear response function for an inhomogeneous system of
electrons in equilibrium with an array of fixed ions is considered. Two
routes to its evaluation for extreme conditions (e.g., warm dense matter)
are considered. The first is from a recently developed short-time kinetic
equation; the second is from time-dependent density functional theory
(tdDFT). The result from the latter approach agrees with that from kinetic
theory in the ``adiabatic approximation", providing support and 
contextual clarity
 for each. Both provide a connection to the
phenomenological Kubo-Greenwood method for calculating transport properties. 
\end{abstract}

\date{August 21, 2018; original May 14, 2018}
\maketitle

\section{Introduction}

\label{sec1}Matter under extreme conditions is of broad current interest,
ranging from applications in theoretical astrophysics (e.g., massive
exo-planets) to new experimental access to such materials \cite{Glenzer16}.
The state conditions include those for which\ many traditional methods of
plasma physics or condensed matter physics fail or become uncontrolled.
However, thermodynamic properties such as pressure, free energy, and
structure are treated well by \textit{ab initio} molecular dynamics (AIMD)
methods \cite{AIMD}, wherein complex electronic states are described by
finite temperature density functional theory (DFT). These methods allow
inclusion of strong coupling, bound and free states, and quantum effects
across a wide range of temperatures and densities. Transport properties and
other dynamical features require an extension of these tools \cite{Dharma17}%
. One approach is a recently developed short-time kinetic equation for time
correlation functions \cite{Dufty17}.{\ I}t subsumes a practical
phenomenology, the Kubo-Greenwood (KG) method \cite{Dufty18,KG}, used for
calculating correlation functions. This approach models the true many-body
Hamiltonian by one for non-interacting particles whose excitations are those
of the equilibrium Kohn-Sham Hamiltonian. The KG method exploits strong
coupling features of equilibrium DFT, extending its advantages to
time-dependent properties.

A second approach is time-dependent density functional theory (tdDFT)
designed to extend the advantages of equilibrium DFT to dynamical properties 
\cite{Marquesbook,Ullrich12,Botti07}. Its formulation and application to
ground state properties is well-developed, but much less so for the finite
temperature extended systems considered here. An extension of van
Leeuwen's fundamental theorem for tdDFT \cite{vanL99}, to mixed states
(ensembles) \cite{Li85,PJ16} is proposed and discussed in Appendix \ref{appA}.
Its application to linear response \cite{Lresp} about an initial
equilibrium state is described in section \ref{sec4}. In particular, it is
shown that the density response function from tdDFT can be expressed in
terms of the KG response function, so its connection to the KG phenomenology
is quite direct. Both the kinetic theory and tdDFT provide means to include
corrections to the KG method. In the ``adiabatic approximation" tdDFT gives
corrections that are equivalent to those from kinetic theory, thereby
establishing a connection between these two quite different approaches.

Here we address three different groups: 1) those focused upon applications
(simulations and experiments) in warm, dense matter, 2) kinetic theory
specialists in many-body physics, and 3) time-dependent density functional
theorists, mainly from atomic and molecular physics. Typically one group
does not follow the literature of the others. The result is loss of insight.
We have tried to make the presentation simple, direct, and self-contained
for value to all three.

\section{Density response and related properties}

\label{sec2}Linear response for systems at initial equilibrium is treated in
most text books on condensed matter physics \cite{McLennan89,Vignale05}. A
recent updated discussion of linear response can be found in reference \cite%
{Stefanucci13}. However, for definition of notation and precise
specification of linear response as used herein, a brief review follows.
Consider a one component system of $N$ identical particles with Hamiltonian $%
H(t)$ 
\begin{equation}
H\left( t\right) =K+U+V(t).  \label{2.1}
\end{equation}%
Here $K$ is the kinetic energy, $U$ is a many-body potential energy among
the particles (more specifically, below this is chosen to be the Coulomb
interactions among electrons and between them and a configuration of fixed
ions), and $V(t)$ is an external time dependent potential (perturbation) of
the form%
\begin{equation}
V(t)=\int d\mathbf{r}v(\mathbf{r},t)\widehat{n}(\mathbf{r}),\hspace{0.25in}%
\widehat{n}(\mathbf{r})=\sum_{i=1}^{N}\delta (\mathbf{r-q}_{i})  \label{2.2}
\end{equation}%
The number density operator is defined in terms of the particle position
operators $\left\{ \mathbf{q}_{i}\right\} $ (a caret is included on $%
\widehat{n}(\mathbf{r})$ in this definition to distinguish the operator from
its state-averaged value $n(\mathbf{r})$ introduced below). The form of the
external potential $v(\mathbf{r},t)$ is unspecified at this point. The state
of the system is given by its density matrix $\rho (t)$. Its evolution is
governed by the Liouville-von Neumann equation, for $t\geq \tau $ 
\begin{equation}
\partial _{t}\rho (t)+i\left[ H(t),\rho (t)\right] =0,  \label{2.3}
\end{equation}%
with some given initial condition $\rho (\tau )$.

Choose the initial state $\rho(\tau)$ to be stationary (equilibrium) under
the unperturbed Hamiltonian 
\begin{equation}
\left[ (K+U),\rho_{eq}\right] =0,\hspace{0.25in}\rho(\tau)\equiv\rho_{eq}
\label{2.4}
\end{equation}
Then the solution to Eq. (\ref{2.3})\ to linear order in the perturbation is 
\begin{equation}
\rho(t)=\rho_{eq}-\int_{\tau}^{t}dt^{\prime}\int d\mathbf{r}v(\mathbf{r}%
,t^{\prime})i\left[ \widehat{n}(\mathbf{r},t^{\prime}-t),\rho_{eq}\right] ,
\label{2.5}
\end{equation}
where the time dependence of the local density operator is%
\begin{equation}
\widehat{n}(\mathbf{r},t)=e^{i\left( K+U\right) t}\widehat{n}(\mathbf{r)}%
e^{-i(K+U)t}.  \label{2.6}
\end{equation}
The equilibrium averaged local density to linear order is therefore 
\begin{equation}
n(\mathbf{r},t\mid v)=n_{eq}(\mathbf{r})+\int_{\tau}^{t}dt^{\prime}\int d%
\mathbf{r}^{\prime}\chi\left( \mathbf{r},\mathbf{r}^{\prime};t-t^{\prime
}\right) v(\mathbf{r}^{\prime},t^{\prime}).  \label{2.7}
\end{equation}
The linear response function $\chi\left( \mathbf{r},\mathbf{r}^{\prime
};t\right) $ is identified as 
\begin{equation}
\chi\left( \mathbf{r},\mathbf{r}^{\prime};t\right) \equiv-i\left\langle %
\left[ \widehat{n}(\mathbf{r},t),\widehat{n}(\mathbf{r}^{\prime})\right]
\right\rangle _{eq},  \label{2.8}
\end{equation}
and the bracket\ with subscript $eq$ denotes the equilibrium average over
the initial state, 
\begin{equation}
\left\langle X\right\rangle _{eq}=Tr\rho_{eq}X.  \label{2.9}
\end{equation}
The cyclic invariance of the trace and stationarity of $\rho_{eq}$ have been
used to obtain the form Eq. (\ref{2.8}).

To be more specific, consider the example of a system of $N_{e}$ electrons
in equilibrium with a distribution of $N_{i}$ fixed ions at the initial time 
$\tau$. The Hamiltonian $H\left( \tau\right) =H_{N_{e}}$ is then%
\begin{equation}
H_{N_{e}}=K+U=K+\frac{1}{2}\sum \limits_{i\neq j=1}^{N_{e}}\frac{e^{2}}{%
\left\vert \mathbf{q}_{i}-\mathbf{q}_{j}\right\vert }+\sum
\limits_{i=1}^{N_{e}}V\left( \mathbf{q}_{i},\left\{ \mathbf{R}\right\}
\right) ,  \label{2.10}
\end{equation}
and the interaction potential for each electron with the ions is%
\begin{equation}
V\left( \mathbf{q}_{i},\left\{ \mathbf{R}\right\} \right) \equiv-\sum
\limits_{j=1}^{N_{i}}\frac{Z_{j}e^{2}}{\left\vert \mathbf{q}_{i}-\mathbf{R}%
_{j}\right\vert }.  \label{2.11}
\end{equation}
Also, for the stationary equilibrium state, choose the grand canonical
ensemble%
\begin{equation}
\rho_{eq,N_{e}}=e^{\beta\Omega}e^{-\beta\left( H_{N_{e}}(\tau)-\mu
N_{e}\right) }\mathcal{S}_{N_{e}},  \label{2.11a}
\end{equation}
where $\mu$ is the chemical potential, $\mathcal{S}_{N_{e}}$ is the $N_{e}$
particle anti-symmetrization operator, and $\Omega$ is the normalization
constant%
\begin{equation}
e^{-\beta\Omega}=\sum_{N_{e}>0}Tr^{(N_{e})}e^{-\beta\left(
H_{N_{e}}(\tau)-\mu N_{e}\right) }\mathcal{S}_{N_{e}}.  \label{2.11b}
\end{equation}
Averages in the grand ensemble are defined by%
\begin{equation}
\left\langle X\right\rangle
_{eq}=\sum_{N_{e}>0}Tr^{(N_{e})}\rho_{eq,N_{e}}X_{N_{e}}.  \label{2.11c}
\end{equation}

\subsection{Relationship to dielectric function and conductivity}

Define the Fourier-transformed response function%
\begin{align}
\widetilde{\chi}\left( \mathbf{k},\mathbf{k}^{\prime};t\right) & =\int d%
\mathbf{r}d\mathbf{r}^{\prime}e^{-i(\mathbf{k\cdot r}+\mathbf{k}^{\prime }%
\mathbf{\cdot r}^{\prime})}\chi\left( \mathbf{r},\mathbf{r}^{\prime
};t\right)  \notag \\
& =-i\left\langle \left[ \widetilde{n}(\mathbf{k},t),\widetilde {n}(\mathbf{k%
}^{\prime})\right] \right\rangle _{eq}  \label{2.12}
\end{align}
where $\widetilde{n}(\mathbf{k})$ is the Fourier transform of the number
operator $\widehat{n}(\mathbf{r})$. A related property is the dielectric
function $\epsilon\left( \mathbf{k},\mathbf{k}^{\prime};t\right) $ defined by%
\begin{equation}
\widetilde{V}(\mathbf{k})\widetilde{\chi}(\mathbf{k,k}^{\prime};t)=\delta _{%
\mathbf{k,-k}^{\prime}}-\epsilon^{-1}(\mathbf{k,k}^{\prime};t).  \label{2.13}
\end{equation}
Here $\widetilde{V}(\mathbf{k})$ is the Fourier transform of the
electron-electron Coulomb potential. If $\epsilon(\mathbf{k,k}^{\prime},t)$
is expanded to leading order in $\widetilde{V}(\mathbf{k})$ the random phase
approximation is obtained 
\begin{equation}
\epsilon(\mathbf{k,k}^{\prime},t)\rightarrow\epsilon_{RPA}(\mathbf{k,k}%
^{\prime};t)=\delta_{\mathbf{k,-k}^{\prime}}+\widetilde{V}(\mathbf{k})%
\widetilde{\chi}^{(0)}(\mathbf{k,k}^{\prime};t),  \label{2.14}
\end{equation}
where $\widetilde{\chi}^{(0)}(\mathbf{k,k}^{\prime};t)$ is the response
function for non-interacting electrons in the presence of the external ions.

Other properties of interest are related to $\widetilde{\chi}\left( \mathbf{k%
},\mathbf{k}^{\prime};t\right) $, or equivalently to $\epsilon \left( 
\mathbf{k},\mathbf{k}^{\prime};t\right) $, by the microscopic number density
conservation law%
\begin{equation}
\partial_{t}\widetilde{n}(\mathbf{k},t)+i\mathbf{k}\cdot\widetilde{\mathbf{j}%
}(\mathbf{k,}t)=0,\hspace{0in}\hspace{0.25in}\widetilde{\mathbf{j}}(\mathbf{k%
})=\sum_{i=1}^{N}\frac{1}{2}\left( e^{-i\mathbf{k\cdot q}_{i}}\mathbf{v}_{i}+%
\mathbf{v}_{i}e^{-i\mathbf{k\cdot q}_{i}}\right) ,  \label{2.15}
\end{equation}
where $\widetilde{\mathbf{j}}(\mathbf{k})$ is the Fourier transformed number
flux operator and $\mathbf{v}_{i}=\mathbf{p}_{i}/m$ is the velocity operator
for particle $i$. The time derivative of $\widetilde{\chi}\left( \mathbf{k},%
\mathbf{k}^{\prime};t\right) $ gives%
\begin{equation}
\partial_{t}\widetilde{\chi}\left( \mathbf{k},\mathbf{k}^{\prime};t\right)
=ik_{\ell}\left\langle i\left[ \widetilde{j}_{\ell}(\mathbf{k},t),%
\widetilde {n}(\mathbf{k}^{\prime})\right] \right\rangle _{eq}.  \label{2.16}
\end{equation}
Use the cyclic property of the trace%
\begin{equation}
\left\langle \left[ \widetilde{j}_{\ell}(\mathbf{k},t),\widetilde {n}(%
\mathbf{k}^{\prime})\right] \right\rangle _{eq}=Tr\left[ \widetilde {n}(%
\mathbf{k}^{\prime}),\rho_{e}\right] \widetilde{j}_{\ell}(\mathbf{k},t),
\label{2.17}
\end{equation}
and the operator identity%
\begin{align}
i\left[ \widetilde{n}(\mathbf{k}^{\prime}),e^{-\beta H(\tau))}\right] &
=-\int_{0}^{\beta}d\lambda e^{\left( \beta-\lambda\right) H(\tau))}i\left[ 
\widetilde{n}(\mathbf{k}^{\prime}),H(\tau)\right] e^{-\lambda H(\tau ))} 
\notag \\
& =-\int_{0}^{\beta}d\lambda e^{\left( \beta-\lambda\right) H(\tau ))}i%
\mathbf{k}^{\prime}\cdot\widetilde{\mathbf{j}}(\mathbf{k}^{\prime
})e^{-\lambda H(\tau))}  \label{2.18}
\end{align}
to get%
\begin{equation}
\partial_{t}\widetilde{\chi}\left( \mathbf{k},\mathbf{k}^{\prime};t\right)
=ik_{m}ik_{\ell}\int_{0}^{\beta}d\lambda\left\langle \widetilde{j}_{m}(%
\mathbf{k}^{\prime},-t+i\lambda)\widetilde{j}_{\ell}(\mathbf{k}%
)\right\rangle _{eq}.  \label{2.19}
\end{equation}
Finally, the Fourier transform in time 
\begin{equation}
\widetilde{\widetilde{\chi}}\left( \mathbf{k},\mathbf{k}^{\prime};\omega%
\right) \equiv\int_{-\infty}^{\infty}dte^{i\omega t}\widetilde{\chi }\left( 
\mathbf{k},\mathbf{k}^{\prime};t\right)  \label{2.20}
\end{equation}
gives%
\begin{align}
\widetilde{\widetilde{\chi}}\left( \mathbf{k},\mathbf{k}^{\prime};\omega%
\right) & =i\frac{k_{m}k_{\ell}}{\omega}\int_{-\infty}^{\infty }dte^{i\omega
t}\int_{0}^{\beta}d\lambda\left\langle \widetilde{j}_{m}(\mathbf{k}%
^{\prime},-t+i\lambda)\widetilde{j}_{\ell}(\mathbf{k})\right\rangle _{eq} 
\notag \\
& =2i\frac{k_{m}k_{\ell}}{\omega e^{2}}\sigma_{m\ell}\left( \mathbf{k},%
\mathbf{k}^{\prime};\omega\right)  \label{2.21}
\end{align}
where the electrical conductivity tensor is%
\begin{equation}
\sigma_{m\ell}\left( \mathbf{k},\mathbf{k}^{\prime};\omega\right) =\frac {1}{%
2}e^{2}\int_{-\infty}^{\infty}dte^{i\omega
t}\int_{0}^{\beta}d\lambda\left\langle \widetilde{j}_{m}(\mathbf{k}%
^{\prime},-t+i\lambda )\widetilde{j}_{\ell}(\mathbf{k})\right\rangle _{eq}.
\label{2.22}
\end{equation}

\section{Kubo-Greenwood method}

The response function (and related equilibrium time correlation functions)
is determined from the Hamiltonian, Eq. (\ref{2.10}), which appears both in
the equilibrium distribution function and the dynamics of $\widehat{n}(%
\mathbf{r},t)$ in Eq. (\ref{2.6}). Its evaluation for the conditions of
interest here involves all the difficulties of the many-body problem for
which standard methods of condensed matter physics or plasma physics are
questionable or intractable. Instead, a phenomenological mean-field model
incorporating strong coupling information from equilibrium DFT commonly is
assumed. The actual Hamiltonian is replaced by 
\begin{equation}
H_{N_{e}}\rightarrow H_{KS}\equiv \sum_{i=1}^{N_{e}}h_{KS}(i),\hspace{0.25in}%
h_{KS}(i)=\frac{p_{i}^{2}}{2m}+v_{KS}\left( \mathbf{q}_{i},\left\{ \mathbf{R}%
\right\} \right) .  \label{2.23}
\end{equation}%
This is a sum of independent Hamiltonians in each of which the effective
single particle potential is the Kohn-Sham potential of equilibrium DFT. It
is determined from the equilibrium free energy functional according to%
\begin{equation}
v_{KS}\left( \mathbf{q}_{i},\left\{ \mathbf{R}\right\} \right) =V\left( 
\mathbf{q}_{1},\left\{ \mathbf{R}\right\} \right) +\frac{\delta F^{(1)}}{%
\delta n\left( \mathbf{q}_{1},\left\{ \mathbf{R}\right\} \right) },
\label{2.24}
\end{equation}%
where $F^{(1)}$ is the excess free energy, beyond the corresponding
non-interacting contribution. It is a functional of the initial equilibrium
density $n\left( \mathbf{q}_{1},\left\{ \mathbf{R}\right\} \right) $. It can
be calculated with good confidence for matter under extreme conditions from
recently developed finite temperature equilibrium DFT methods \cite{DFT}.
The approximation Eq. (\ref{2.23}) is known as the Kubo-Greenwood method.
Since it invokes a system of non-interacting particles, the response
function can be calculated exactly, for a given $F^{(1)}$ and configuration
of the ions $\left\{ \mathbf{R}\right\} $, in terms of the eigenfunctions
and eigenvalues of $h_{KS}$ \cite{KG}.

The origin and basis for the Kubo-Greenwood method is not clear beyond the
hope that the reasonably accurate description of interactions for
thermodynamic properties from equilibrium DFT approximations might extend to
the dynamics as well. A major objective of the present work is to provide a
more convincing rationalization for the replacement shown in (\ref{2.23}).

\section{Short time kinetic theory}

\label{sec3}The density response function can be written in the equivalent
form%
\begin{align}
\chi \left( \mathbf{r},\mathbf{r}^{\prime };t\right) & =i\left\langle 
\widehat{n}(\mathbf{r}^{\prime })\widehat{n}(\mathbf{r},t)\right\rangle
_{eq}-i\left\langle \widehat{n}(\mathbf{r},t)\widehat{n}(\mathbf{r}^{\prime
})\right\rangle _{eq}  \notag \\
& =i\left( C\left( \mathbf{r},\mathbf{r}^{\prime };t\right) -C\left( \mathbf{%
r},\mathbf{r}^{\prime };t+i\beta \right) \right) .  \label{3.1a}
\end{align}%
Here $C\left( \mathbf{r},\mathbf{r}^{\prime },t\right) $ is the time
correlation function 
\begin{equation}
C\left( \mathbf{r},\mathbf{r}^{\prime };t\right) =\left\langle \widehat{n}(%
\mathbf{r}^{\prime })\widehat{n}(\mathbf{r},t)\right\rangle _{eq}.
\label{3.2}
\end{equation}%
To obtain the second line of (\ref{3.1a}) the cyclic invariance of the trace
has been used. Following the formal kinetic theory of reference \cite%
{Boercker81}, the correlation function can be written as an average over the
single electron subspace%
\begin{equation}
C\left( \mathbf{r},\mathbf{r}^{\prime };t\right) =Tr_{1}\delta \left( 
\mathbf{r}-\mathbf{q}_{1}\right) \psi \left( 1,\mathbf{r}^{\prime };t\right)
,  \label{3.3}
\end{equation}%
Here $Tr_{1}$\ denotes a trace in the single particle Hilbert space, and the
single particle operator $\psi \left( 1,\mathbf{r};t\right) $\ is averaging
over all other degrees of freedom (analogous to a one-particle reduced
density matrix but representing the correlation function). It obeys the
formally exact kinetic, equation%
\begin{equation}
\left( \partial _{t}+B\left( 1\right) \right) \psi \left( 1,\mathbf{r}%
^{\prime };t\right) =\int_{0}^{t}dt^{\prime }M\left( 1;\ t^{\prime }\right)
\psi \left( 1,\mathbf{r}^{\prime };t-t^{\prime }\right) \,,  \label{3.4}
\end{equation}%
where $B$ and $M$ are super operators that map the single particle Hilbert
space operators onto other single particle operators. For the present, the
detailed formal definitions for $B\left( 1\right) $ and $M\left( 1;t\right) $
are not needed, beyond the facts that $B\left( 1\right) $ is time
independent and $M\left( 1;t\right) $ is non-singular at $t=0$. This means
that the exact short time form for the kinetic theory is 
\begin{equation}
\left( \partial _{t}+B\left( 1\right) \right) \psi \left( 1,\mathbf{r}%
^{\prime };t\right) =0,\hspace{0.25in}t\rightarrow 0.  \label{3.5}
\end{equation}%
Use of this form for $t>0$ constitutes the Markov approximation, whereby the
generator of the time dependence does not depend on time. Such an
approximation does not involve 
any explicit limitation on coupling
strength or other small parameter conditions. Hence it is a good candidate
for materials under extreme conditions.

The correlation function $C\left( \mathbf{r},\mathbf{r}^{\prime };t\right) $
calculated using this short time kinetic theory is obtained by integrating (%
\ref{3.5}) 
\begin{equation}
C\left( \mathbf{r},\mathbf{r}^{\prime };t\right) =Tr_{1}\delta \left( 
\mathbf{r}-\mathbf{q}_{1}\right) e^{-Bt}\psi \left( 1,\mathbf{r}^{\prime
},0\right) ,  \label{3.6}
\end{equation}%
and the corresponding response function from (\ref{3.1a}) is%
\begin{equation}
\chi \left( \mathbf{r},\mathbf{r}^{\prime };t\right) =iTr_{1}\delta \left( 
\mathbf{r}-\mathbf{q}_{1}\right) e^{-Bt}\left( 1-e^{-i\beta B}\right) \psi
\left( 1,\mathbf{r}^{\prime };0\right) .  \label{3.7}
\end{equation}%
This can be simplified using the exact initial value for $\chi \left( 
\mathbf{r},\mathbf{r}^{\prime };t\right) $ calculated directly from its
definition Eq. (\ref{2.12})%
\begin{equation}
\chi \left( \mathbf{r},\mathbf{r}^{\prime };0\right) =iTr_{1}\delta \left( 
\mathbf{r}-\mathbf{q}_{1}\right) \left[ f^{(1)}(1),\delta \left( \mathbf{r}%
^{\prime }-\mathbf{q}_{1}\right) \right] .  \label{3.8}
\end{equation}%
where $f^{(1)}(1)$ is the single-electron equilibrium distribution operator%
\begin{equation}
f^{(1)}(1)=\sum_{N_{e}\geq 2}N_{e}Tr_{2..N_{e}}\rho _{eN_{e}},  \label{3.9}
\end{equation}%
and $\rho _{eN_{e}}$ is the grand canonical equilibrium state of (\ref{2.11a}%
)\ and the trace $Tr_{2..N_{e}}$\ is taken over all degrees of freedom
except index $1$. This determines $\psi \left( 1,\mathbf{r};0\right) $ in
terms of $B$%
\begin{equation}
\left( 1-e^{-i\beta B}\right) \psi \left( 1,\mathbf{r}^{\prime };0\right) =%
\left[ f^{(1)}(1),\delta \left( \mathbf{r}^{\prime }-\mathbf{q}_{1}\right) %
\right]   \label{3.9a}
\end{equation}%
to give the final short-time kinetic theory result for the response function%
\begin{equation}
\chi \left( \mathbf{r},\mathbf{r}^{\prime };t\right) =Tr_{1}\delta \left( 
\mathbf{r}-\mathbf{q}_{1}\right) \phi \left( 1,\mathbf{r}^{\prime };t\right)
,  \label{3.10}
\end{equation}

\begin{equation}
\left( \partial _{t}+B\left( 1\right) \right) \phi \left( 1,\mathbf{r}%
^{\prime };t\right) =0,\hspace{0.25in}\phi \left( 1,\mathbf{r}^{\prime
};0\right) =i\left[ f^{(1)}(1),\delta \left( \mathbf{r}^{\prime }-\mathbf{q}%
_{1}\right) \right] .  \label{3.11}
\end{equation}

As an example, the calculation of $B\left( 1\right) $ in the weak coupling
limit is given in reference \cite{Boercker81}, leading to $B\left( 1\right) $
for the random phase approximation linear kinetic equation%
\begin{align}
B\left( 1\right) \phi \left( 1,\mathbf{r}^{\prime };t\right) & \rightarrow
i[\left( \frac{p_{1}^{2}}{2m}+V\left( \mathbf{q}_{1},\left\{ \mathbf{R}%
\right\} \right) \right) ,\phi \left( 1,\mathbf{r}^{\prime };t\right) ] 
\notag \\
& +Tr_{2}i[V_{ee}(12),f^{(2)}(12)f^{-1}(1)\phi \left( 1,\mathbf{r}^{\prime
};t\right) ]  \notag \\
& +Tr_{2}i[V_{ee}(12),f^{(2)}(12)f^{-1}(2)\phi \left( 2,\mathbf{r}^{\prime
};t\right) ].  \label{3.12}
\end{align}%
Here, $f(1)$ and $f^{(2)}(12)$ are the non-interacting one- and two-particle
reduced density operators, including exchange, and $V_{ee}(12)=e^{2}/\mid {%
\mathbf{q}}_{1}-{\mathbf{q}}_{2}\mid $. The second term on the right of Eq. (%
\ref{3.12}) represents the Hartree-Fock additions to the single-particle
energies, while the third term gives the RPA screening. More generally, to
include strong coupling effects, $B\left( 1\right) $ has been expressed
exactly in terms of the one-, two-, and three-particle equilibrium reduced
density matrices for the interacting system \cite{Boercker81}. However, a
more practical representation has been obtained only in the semi-classical
limit. 
That invokes a classical representation for the electrons
with short-distance regularization of the Coulomb potentials for
electron-electron and electron-ion interactions to account for quantum
diffraction and exchange effects. In that case $B\left( 1\right) $ can be
calculated exactly without any limitations on the coupling strength between
electrons or electrons and ions \cite{Dufty17}, and its quantization
performed \emph{a posteriori} (see section V of reference \cite{Dufty17}).
The result again is in the form of the random phase approximation but with
the ion-electron and electron-electron potentials renormalized for strong
coupling 
\begin{align}
B\left( 1\right) \phi \left( 1,\mathbf{r}^{\prime };t\right) & =i[\left( 
\frac{p_{1}^{2}}{2m}+\mathcal{V}\left( \mathbf{q}_{1},\left\{ \mathbf{R}%
\right\} \right) \right) ,\phi \left( 1,\mathbf{r}^{\prime };t\right) ] 
\notag \\
& +Tr_{2}i[\mathcal{V}_{ee}(12),f^{(1)}(1)\phi \left( 2,\mathbf{r}^{\prime
};t\right) ].  \label{3.13}
\end{align}%
with%
\begin{equation}
\mathcal{V}\left( \mathbf{q}_{1},\left\{ \mathbf{R}\right\} \right) =-\frac{%
\delta F^{(0)}(\beta \mid n)}{\delta n\left( \mathbf{q}_{1},\left\{ \mathbf{R%
}\right\} \right) }  \label{3.14}
\end{equation}%
\begin{equation}
\mathcal{V}_{ee}(12)=\mathcal{V}_{ee}(\mathbf{q}_{1},\mathbf{q}_{2})=\frac{%
\delta ^{2}F^{(1)}(\beta ,\left\{ \mathbf{R}\right\} \mid n)}{\delta n\left( 
\mathbf{q}_{1},\left\{ \mathbf{R}\right\} \right) \delta n\left( \mathbf{q}%
_{2},\left\{ \mathbf{R}\right\} \right) }  \label{3.15}
\end{equation}%
Note that these are evaluated at the density of the equilibrium reference
state. The free energy for the system $F=F^{(0)}+F^{(1)}$ has been separated
into its non-interacting and excess parts. The non-interacting part is
related to the Kohn-Sham potential of equilibrium DFT $\delta F^{(0)}(\beta
\mid n)/\delta n\left( \mathbf{r},\left\{ \mathbf{R}\right\} \right) \equiv
\mu -v_{KS}(\mathbf{r},\left\{ \mathbf{R}\right\} )$ so%
\begin{equation}
\mathcal{V}\left( \mathbf{r},\left\{ \mathbf{R}\right\} \right) =v_{KS}(%
\mathbf{r},\left\{ \mathbf{R}\right\} )-\mu .  \label{3.16}
\end{equation}%
The chemical potential $\mu $ does not contribute to the first term on the
right side of Eq. (\ref{3.13}), so this becomes the commutator with the
Kohn-Sham Hamiltonian of Eq. (\ref{2.23}).

The short time kinetic theory Eq. (\ref{3.11}) now becomes%
\begin{equation}
\partial_{t}\phi\left( 1,\mathbf{r}^{\prime};t\right)
+i[h_{KS}(1),\phi\left( 1,\mathbf{r}^{\prime};t\right) ]=-Tr_{2}i[\mathcal{V}%
_{ee}(12),f^{(1)}(1)\phi\left( 2,\mathbf{r}^{\prime};t\right) ].
\label{3.17}
\end{equation}
The left side of this equation describes independent particle dynamics
generated by the Kohn-Sham Hamiltonian%
\begin{equation}
h_{KS}(1)=\frac{p_{1}^{2}}{2m}+\mathcal{V}\left( \mathbf{q}_{1},\left\{ 
\mathbf{R}\right\} \right) .  \label{3.17a}
\end{equation}
This is precisely the generator for the dynamics of the KG method. Indeed,
if $\mathcal{V}_{ee}(12)$ is set equal to zero on the right side of Eq. (\ref%
{3.17}) the resulting kinetic theory is equivalent to that method. The more
general short-time kinetic theory therefore provides some context for the KG
method, and shows that renormalized RPA screening by the electrons is
neglected in that method. Further comment on this connection is given in
section \ref{sec5}.

The short-time kinetic equation solution as given in Appendix \ref{appB}
determines the density response function. The result is given by the linear
integral equation%
\begin{equation}
\chi(\mathbf{r,r}^{\prime},t)=\chi_{KG}\left( \mathbf{r},\mathbf{r}^{\prime
};t\right) +\int_{0}^{t}dt^{\prime}\int d\mathbf{r}_{1}d\mathbf{r}%
_{2}\chi_{KG}\left( \mathbf{r},\mathbf{r}_{1};t-t^{\prime}\right) \mathcal{V}%
_{ee}\left( \mathbf{r}_{1},\mathbf{r}_{2}\right) \chi\left( \mathbf{r}_{2},%
\mathbf{r}^{\prime};t^{\prime}\right) .  \label{3.18}
\end{equation}
Here $\chi_{KG}$ is the response function calculated with the Kohn-Sham
Hamiltonian Eq. (\ref{2.23}), i.e.\ that from the KG method.

\section{Time-dependent density functional theory}

\label{sec4}Time-dependent density functional theory is a well developed
tool within dynamic electronic structure methods with a wide range of
applications to problems in atomic, molecular, and extended systems in
physics, chemistry, and materials science \cite{Marquesbook,Ullrich12}.
Typically such applications are pure state dynamics. Formulation and
application of tdDFT to the mixed state ensembles at finite temperatures of
interest here is more limited \cite{Li85,PJ16}. However, an interesting
calculation of x-ray Thomson scattering for warm, dense matter conditions
has been reported recently \cite{xray}. Central to that formulation are the
consequences of van Leeuwen's theorem on existence and uniqueness of a
time-dependent density representation \cite{vanL99,Marquesbook,Ullrich12}. 
For completeness, an extension of van Leeuwen's theorem for general
mixed states, including those of thermal equilibrium, is proposed in
Appendix \ref{appA}. The argument for this extension assumes physically
reasonable behavior (e.g. invertibility, analyticity) to make the point
without addressing mathematical difficulties well-known in the pure state
case \cite{Ruggenthaler}.

Consider again the system of electrons in a charge-neutral background of a
given ion configuration at equilibrium. The Hamiltonian is that of (\ref%
{2.10}) and the initial state at time $\tau $ is given by (\ref{2.11a}).
Under a time dependent perturbation $V(t)=\int d\mathbf{r}v(\mathbf{r},t)%
\widehat{n}(\mathbf{r})$, its average density for $t\geq \tau $ is denoted
by $n(\mathbf{r},t\mid v)$. A consequence of van Leeuwen's theorem is the
existence of a unique external perturbation $V_{0}(t)=\int d\mathbf{r}v_{0}(%
\mathbf{r},t)\widehat{n}(\mathbf{r})$ such that the corresponding system
without electron-electron interactions produces the same average time
dependent density 
\begin{equation}
n_{0}(\mathbf{r},t\mid v_{0})=n(\mathbf{r},t\mid v).  \label{4.1}
\end{equation}%
where $n_{0}(\mathbf{r},t\mid v_{0})$ is the average density without
electron-electron interactions, in the external potential $V_{0}(t)$. By
continuity, it is expected that $v_{0}\rightarrow 0$ as $v\rightarrow 0$ and
therefore that this equivalence of densities is preserved to linear order in
the two perturbations. Then, repeating the linear response analysis of
Section \ref{sec2} leads to the equivalence in the initial state%
\begin{equation}
n_{0}(\mathbf{r},\tau \mid v_{0})=n(\mathbf{r},\tau \mid v),  \label{4.1a}
\end{equation}%
and at later times%
\begin{equation}
\int_{\tau }^{t}dt^{\prime }\int d\mathbf{r}^{\prime }\chi \left( \mathbf{r},%
\mathbf{r}^{\prime };t-t^{\prime }\right) \delta v(\mathbf{r}^{\prime
},t^{\prime })=\int_{\tau }^{t}dt^{\prime }\int d\mathbf{r}^{\prime }\chi
_{0}\left( \mathbf{r},\mathbf{r}^{\prime };t-t^{\prime }\right) \delta v_{0}(%
\mathbf{r}^{\prime },t^{\prime })  \label{4.2}
\end{equation}%
Here, $\chi _{0}\left( \mathbf{r},\mathbf{r}^{\prime };t\right) $ is the
response function for the initial non-interacting system.

Equation (\ref{4.1a}) is a first condition of van Leeuwen's theorem, that
the initial densities should be the same. Furthermore, since the unperturbed
states are equilibrium, it follows from equilibrium DFT that the external
potential for the non-interacting system at $t=\tau$ is the Kohn-Sham
potential as a functional of this initial density%
\begin{equation}
v_{0}(\mathbf{r},\tau)=v_{KS}(\mathbf{r}\mid n_{e}).  \label{4.3}
\end{equation}
Consequently, $\chi_{0}\left( \mathbf{r},\mathbf{r}^{\prime};t\right) $ is
the response function defined by the Kohn-Sham Hamiltonian, both for its
equilibrium average and for the generator of its time dependence; this is
then the Kubo-Greenwood response function%
\begin{equation}
\chi_{0}\left( \mathbf{r},\mathbf{r}^{\prime};t\right) =\chi_{KG}\left( 
\mathbf{r},\mathbf{r}^{\prime};t\right) .  \label{4.4}
\end{equation}
Since this is a non-interacting system, it can be evaluated exactly in terms
of the eigenvalues and eigenfunctions of the Kohn-Sham Hamiltonian.

It is a remarkable consequence of van Leeuwen's theorem that the equivalence
of the densities allows the more complex interacting system response
function to be related to this simpler non-interacting response function.
More explicitly, from (\ref{4.2}) 
\begin{align}
\chi\left( \mathbf{r},\mathbf{r}^{\prime};t-t^{\prime}\right) &
=\int_{\tau}^{t}dt^{\prime\prime}\int d\mathbf{r}^{\prime\prime}\chi
_{KG}\left( \mathbf{r},\mathbf{r}^{\prime\prime};t-t^{\prime\prime}\right) 
\frac{\delta v_{0}(\mathbf{r}^{\prime\prime},t^{\prime\prime})}{\delta v(%
\mathbf{r}^{\prime},t^{\prime})}  \notag \\
& =\chi_{KG}\left( \mathbf{r},\mathbf{r}^{\prime};t-t^{\prime}\right)
+\int_{\tau}^{t}dt^{\prime\prime}\int d\mathbf{r}^{\prime\prime}\chi
_{KG}\left( \mathbf{r},\mathbf{r}^{\prime\prime};t-t^{\prime\prime}\right) 
\frac{\delta\Delta v_{0}(\mathbf{r}^{\prime\prime},t^{\prime\prime}\mid n)}{%
\delta v(\mathbf{r}^{\prime},t^{\prime})}  \label{4.5}
\end{align}
In the second equality the unknown potential $v_{0}(\mathbf{r},t)$ has been
written as the given potential plus the ``excess potential" $\Delta v_{0}$%
\begin{equation}
v_{0}(\mathbf{r},t)\equiv v(\mathbf{r},t\mid n)+\Delta v_{0}(\mathbf{r}%
,t\mid n),  \label{4.6}
\end{equation}
The notation recognizes that the one-to-one relationship of $n_{0}(\mathbf{r}%
,t\mid v_{0})$ to the potential $v_{0}(\mathbf{r},t)$ implies it can be
inverted to give 
\begin{equation}
\Delta v_{0}(\mathbf{r},t)=\Delta v_{0}(\mathbf{r},t\mid n_{0})=\Delta v_{0}(%
\mathbf{r},t\mid n).  \label{4.7}
\end{equation}
(The first equality states the one-to-one relationship of the
non-interacting potential to the non-interacting density. The second
equality states that the non-interacting and interacting densities are the
same, a consequence of the central property of the KS potential.) Then by
the chain rule 
\begin{align}
\frac{\delta\Delta v_{0}(\mathbf{r},t\mid n)}{\delta v(\mathbf{r}^{\prime
},t^{\prime})} & =\int_{\tau}^{t}dt_{1}\int d\mathbf{r}_{1}\frac{%
\delta\Delta v_{0}(\mathbf{r},t\mid n)}{\delta n(\mathbf{r}_{1},t_{1})}\frac{%
\delta n(\mathbf{r}_{1},t_{1})}{\delta v(\mathbf{r}^{\prime},t^{\prime})} 
\notag \\
& =\int_{\tau}^{t}dt_{1}\int d\mathbf{r}_{1}\frac{\delta\Delta v_{0}(\mathbf{%
r},t\mid n)}{\delta n(\mathbf{r}_{1},t_{1})}\chi\left( \mathbf{r}_{1},%
\mathbf{r}^{\prime};t_{1}-t^{\prime}\right)\,.  \label{4.8}
\end{align}
The final form for the relationship of $\chi$ to $\chi_{KG}$ becomes,
setting $t^{\prime}=0$ in (\ref{4.5}) for simplicity of notation, 
\begin{align}
\chi\left( \mathbf{r},\mathbf{r}^{\prime};t\right) & =\chi_{KG}\left( 
\mathbf{r},\mathbf{r}^{\prime};t\right)
+\int_{\tau}^{t}dt^{\prime\prime}\int d\mathbf{r}^{\prime\prime}\chi_{KG}%
\left( \mathbf{r},\mathbf{r}^{\prime \prime};t-t^{\prime\prime}\right)
\int_{\tau}^{t}dt_{1}\int d\mathbf{r}_{1}  \notag \\
& \times\frac{\delta\Delta v_{0}(\mathbf{r}^{\prime\prime},t^{\prime\prime
}\mid n)}{\delta n(\mathbf{r}_{1},t_{1})}\chi\left( \mathbf{r}_{1},\mathbf{r}%
^{\prime};t_{1}\right) .  \label{4.9}
\end{align}

The result Eq. (\ref{4.9}) is formally exact and is simply a restatement of
the consequence of van Leeuwen's theorem Eq. (\ref{4.1}) to first order in
the perturbing potentials. Interestingly, the appearance of the KG response
function $\chi _{KG}$ also is a consequence of this theorem which requires
that the initial density of the non-interacting and interacting systems
should be the same. For the initial equilibrium state that implies Eq. (\ref%
{4.3}), and hence the Hamiltonian for the non-interacting system is the sum
of Kohn-Sham single particle Hamiltonians. This provides an important
connection with the KG method and a clarification of its logical context.

The excess potential $\Delta v_{0}(\mathbf{r},t\mid n)$ remains unknown. While van Leeuwen's theorem provides its existence, the theorem does
not provide the explicit functional dependence of 
$\Delta v_{0}(\mathbf{r}%
,t\mid n)$ upon $n$.  
However, this dependence is known initially
from Eq. (\ref{4.3}). A plausible approximation is to assume this functional
form persists and that its evolution occurs entirely through the density%
\begin{equation}
v_{0}(\mathbf{r},t\mid n)\sim v_{KS}(\mathbf{r}\mid n\left( t\right) ),
\label{4.10}
\end{equation}%
i.e., the functional form is slowly varying and the dominant change is due
to that of its argument. This is referred to as the \textquotedblleft
adiabatic approximation\textquotedblright\ of tdDFT \cite%
{ALDA,Marquesbook,Ullrich12}. With this approximation%
\begin{align}
\frac{\delta \Delta v_{0}(\mathbf{r}^{\prime \prime },t^{\prime \prime }\mid
n)}{\delta n(\mathbf{r}_{1},t_{1})}& \rightarrow \frac{\delta \Delta v_{KS}(%
\mathbf{r}^{\prime \prime }\mid n\left( t^{\prime \prime }\right) )}{\delta
n(\mathbf{r}_{1},t_{1})}=\frac{\delta ^{2}F^{(1)}[n\left( t^{\prime \prime
}\right) ]}{\delta n(\mathbf{r}_{1},t_{1})\delta n(\mathbf{r}^{\prime \prime
},t^{\prime \prime })}  \notag \\
& =\delta \left( t_{1}-t^{\prime \prime }\right) \frac{\delta
^{2}F^{(1)}[n\left( t^{\prime \prime }\right) ]}{\delta n(\mathbf{r}%
_{1},t^{\prime \prime })\delta n(\mathbf{r}^{\prime \prime },t^{\prime
\prime })}  \label{4.11}
\end{align}%
and the response function Eq. (\ref{4.9}) becomes 
\begin{align}
\chi \left( \mathbf{r},\mathbf{r}^{\prime };t\right) & =\chi _{KG}\left( 
\mathbf{r},\mathbf{r}^{\prime };t\right) +\int_{\tau }^{t}dt^{\prime \prime
}\int d\mathbf{r}^{\prime \prime }\chi _{KG}\left( \mathbf{r},\mathbf{r}%
^{\prime \prime };t-t^{\prime \prime }\right) \int d\mathbf{r}_{1}  \notag \\
& \times \frac{\delta ^{2}F^{(1)}[n_{e}]}{\delta n_{e}(\mathbf{r}_{1})\delta
n_{e}(\mathbf{r}^{\prime \prime })}\chi \left( \mathbf{r}_{1},\mathbf{r}%
^{\prime };t^{\prime \prime }\right) .  \label{4.12}
\end{align}%
The density $n\left( t^{\prime \prime }\right) $ is given by Eq. (\ref{2.7})
so within this context of linear response it has been replaced on the right
side of Eq. (\ref{4.12}) by the reference state density $n(\mathbf{r}%
,t^{\prime \prime })\rightarrow n(\mathbf{r},\tau )=n_{e}(\mathbf{r}).$ Note
that the adiabatic approximation does not make any reference to limitations
on the electron-electron or electron-ion coupling, hence is an appropriate
description for matter under extreme conditions. Remarkably, it is seen that
this result from tdDFT is the same as Eq. (\ref{3.18}) from the Markov
kinetic theory.

\section{Discussion}

\label{sec5}The presentation here is complementary to that of reference \cite%
{PJ16}. The version of the van Leeuwen theorem in that reference is less
comprehensive than that of the Appendix here, in that it refers only to
uniqueness (not existence) and only within the context of linear response.
On the other hand, the objectives of that reference were to set the stage
for improvements of the adiabatic approximation while here the interest is
in making connections to other methods within that approximation. 
Specifically, the objective of the treatment presented here has been to
describe the density response function for matter under extreme conditions.
This means conditions of strong Coulomb coupling with both free and bound
electronic configurations. The detailed form of the interaction potential $U$
in Eq. (\ref{2.1}) is not important for the analysis presented here, but an
important case is electrons in the presence of a given ionic configuration.
Two methods for calculation have been presented, one based in kinetic theory
and the other in tdDFT. In both cases the results are expressed in terms of
effective interactions that can be obtained from well-developed methods of
equilibrium DFT, i.e. functional derivatives of the free energy \cite{DFT}. Interestingly, approximations to the kinetic theory (short-time
Markov limit) and to tdDFT (adiabatic approximation) are found to give
equivalent results, Eq. (\ref{3.18}) or Eq. (\ref{4.12}). Neither of these
approximations compromises extreme conditions (although some physical
processes are excluded) and hence the result is a good candidate for
predictive properties. It has a form similar to that of the RPA. However,
the non-interacting response function in RPA is replaced by $\chi _{KG}$
which is determined from non-interacting Kohn-Sham single particle
Hamiltonians. In this way the electron-ion interaction is described by $%
v_{KS}$ rather than the bare ion-electron Coulomb potential. Similarly, the
RPA screening due to the electron-electron Coulomb potential is replaced by
that due to the renormalized potential $\mathcal{V}_{ee}$ of Eq. (\ref{3.15}%
).

The excluded physical processes alluded to above are electron-electron
collisional effects. The RPA structure includes mean-field electron-electron
screening but not electron-electron scattering. In contrast, for the example
above of electrons in the external field of ions, the electron-ion
"collisions" are treated in detail by the dynamics of the Kohn-Sham
Hamiltonian determining $\chi_{KS}$. In addition to neglecting these
electron-electron collisions, the Kubo-Greenwood method is recovered only if
the screening effects found here are negligible as well. Thus, an important
outcome of the analysis here is to show how the Kubo-Greenwood method
appears as an important component of the response function calculation, and
also to demonstrate its context - neglect of electron-electron scattering
and dynamical screening.

Another interesting outcome is the equivalence of the response function from
the short-time kinetic theory and from tdDFT in the adiabatic approximation.
In hindsight this is perhaps to be expected since each becomes exact in the
short time limit (e.g., compare Eqs. (\ref{4.3}) and (\ref{4.10})). This
close connection provides some potential to explore approximations in tdDFT
beyond the adiabatic approximation. For example the collision operator, $M$,
of the exact kinetic equation, Eq. (\ref{3.4}), has been studied in some
detail \cite{Boercker81} and may provide a route for corresponding
improvements of tdDFT applications.

\section{Acknowledgments}

\label{sec6}This research has been supported by US DOE Grant DE-SC0002139.

\bigskip

\appendix

\section{A proposed generalization of van Leeuwen's theorem for mixed states}

\label{appA} 
Based on the extensive studies of van Leeuwen's theorem
for pure states, it is reasonable to suppose a corresponding theorem applies
for mixed states as well. A complete characterization of the states and
necessary conditions is not the objective here. Instead, a constructive
argument, at physically plausible levels of rigor, is given to demonstrate
van Leeuwen's theorem in the rather general context of ensembles or density
matrices as states for the system. We do not revisit the multiple issues of
a mathematically complete investigation encountered for pure states over the
past two decades. Thus we assume properties such as invertibility,
analyticity, etc. are satisfied as required. Readers interested in those
issues should consult the recent review for pure states by Ruggenthaler et
al. \cite{Ruggenthaler}. A more complete justification of the result
presented here is under consideration for a future publication.

Consider two systems characterized by the Hamiltonians $H\left( t\right) $
and $H_{1}\left( t\right) $%
\begin{equation}
H\left( t\right) =K+U+V(t),\hspace{0.25in}H_{1}\left( t\right)
=K+U_{1}+V_{1}(t).  \label{a.1}
\end{equation}%
Here, $K$ denotes the kinetic energy, $U$ and $U_{1}$ are general many-body
potentials, and $V$ and $V_{1}$ are sums of single particle potentials 
\begin{equation}
V(t)=\int d\mathbf{r}v(\mathbf{r},t)\widehat{n}(\mathbf{r}),\hspace{0.25in}%
V_{1}(t)=\int d\mathbf{r}v_{1}(\mathbf{r},t)\widehat{n}(\mathbf{r}).
\label{a.2}
\end{equation}%
The number density operator $\widehat{n}(\mathbf{r})$ is given by (\ref{2.2}%
). The expectation value of some observable corresponding to an operator $X$
is%
\begin{equation}
\left\langle X\right\rangle =Tr\rho X,\hspace{0.25in}Tr\rho =1.  \label{a.3}
\end{equation}%
The trace is taken over an arbitrary complete set of states defining the
Hilbert space considered.The state of the system is represented by the
positive, semi-definite Hermitian operator $\rho $ normalized to unity. If
it is a projection operator onto a single vector in the Hilbert space it is
referred to as a pure state. Otherwise, it is a mixed state. The
corresponding quantities for the second system are the same but
distinguished by a subscript $1$.

The time-dependence of a state $\rho\left( t\right) $ is given by the
Liouville - von Neumann equation%
\begin{equation}
\partial_{t}\rho(t)=-i\left[ H\left( t\right) ,\rho(t)\right] ,\hspace{0.25in%
}\rho(t=0)=\rho.  \label{a.4}
\end{equation}
where without loss of generality the initial time is taken to be $t=0$.
Accordingly, the average number densities for the two systems are%
\begin{equation}
n(\mathbf{r},t\mid v)=Tr\rho\left( t\right) \widehat{n}(\mathbf{r}%
)\equiv\left\langle \widehat{n}(\mathbf{r});t\right\rangle ,\hspace{0.25in}%
n_{1}(\mathbf{r},t\mid v_{1})=Tr\rho_{1}\left( t\right) \widehat {n}(\mathbf{%
r})\equiv\left\langle \widehat{n}(\mathbf{r});t\right\rangle _{1}
\label{a.5}
\end{equation}
The notation $n(\mathbf{r},t\mid v)$ indicates that the density is a
space-time functional of $v(\mathbf{r},t)$. Also the subscript on the
bracket $\left\langle \widehat{n}(\mathbf{r});t\right\rangle _{1}$ indicates
an average over $\rho_{1}\left( t\right) $ whose dynamics is generated by $%
H_{1}\left( t\right) $. The objective here is to show that for a given $n(%
\mathbf{r},t\mid v)$ there exists a unique $v_{1}(\mathbf{r},t)$ such that $%
n_{1}(\mathbf{r},t\mid v_{1})=n(\mathbf{r},t\mid v)$. The demonstration is
based on direct construction of $v_{1}(\mathbf{r},t)$ from all of its
initial time derivatives under the assumption that the density is analytic
at $t=0$ and upon some domain of non-zero radius \cite%
{vanL99,Marquesbook,Ullrich12}.

Assume there exists a $v_{1}(\mathbf{r},t)$ such that the densities are equal%
\begin{equation}
n(\mathbf{r},t\mid v)=n_{1}(\mathbf{r},t\mid v_{1}),  \label{a.6}
\end{equation}
which gives the formal definition of $v_{1}(\mathbf{r},t)$. The right side
evolves according to the von Neumann equation 
\begin{equation}
\partial_{t}\rho_{1}(t)=-i\left[ H_{1}\left( t\right) ,\rho_{1}(t)\right] ,
\label{a.7}
\end{equation}
or equivalently%
\begin{equation}
\rho_{1}(t)=\rho_{1}(0)-\int_{0}^{t}dt^{\prime}i\left[ H_{1}\left(
t^{\prime}\right) ,\rho_{1}(t^{\prime})\right] .  \label{a.8}
\end{equation}
Then Eq. (\ref{a.6}) becomes%
\begin{align}
n(\mathbf{r},t & \mid v)=\left\langle \widehat{n}(\mathbf{r});0\right\rangle
_{1}-i\int_{0}^{t}dt^{\prime}Tr \left[ H_{1}\left( t^{\prime}\right)
,\rho_{1}(t^{\prime})\right] \widehat{n}(\mathbf{r})  \notag \\
& =\left\langle \widehat{n}(\mathbf{r});0\right\rangle
_{1}+\int_{0}^{t}dt^{\prime}Tr\left\langle i\left[ H_{1}\left(
t^{\prime}\right) ,\widehat{n}(\mathbf{r})\right] ;t^{\prime}\right\rangle
_{1},  \label{a.9}
\end{align}
where the second line follows from the cyclic invariance of the trace. A
further iteration of Eq. (\ref{a.8}) gives 
\begin{align}
n(\mathbf{r},t & \mid v)=\left\langle \widehat{n}(\mathbf{r});0\right\rangle
_{1}+\int_{0}^{t}dt^{\prime}i\left\langle \left[ H_{1}\left( t^{\prime
}\right) ,\widehat{n}(\mathbf{r})\right] ;0\right\rangle _{1}  \notag \\
& +\left( i\right)
^{2}\int_{0}^{t}dt^{\prime}\int_{0}^{t^{\prime}}dt^{\prime\prime}\left%
\langle \left[ H_{1}\left( t^{\prime\prime}\right) ,\left[ H_{1}\left(
t^{\prime}\right) ,\widehat{n}(\mathbf{r})\right] \right] ;t^{\prime\prime}%
\right\rangle _{1}  \label{a.10}
\end{align}
This is still exact.The right side is a functional of $v_{1}(\mathbf{r},t)$
and hence gives its formal definition in terms of the given density $n(%
\mathbf{r},t\mid v)$. Suppose the latter is analytic at $t=0$ so that its
derivatives exist at arbitrary order. Then Eq. (\ref{2.10}) can be expanded
in powers of $t$ and its coefficients of each term identified. A first
condition is that the initial state $\rho_{1}$ must deliver the same density
as $\rho$%
\begin{equation}
n(\mathbf{r},t\mid v)=\left\langle \widehat{n}(\mathbf{r});0\right\rangle
_{1}=Tr\rho_{1}\widehat{n}(\mathbf{r}),  \label{a.10a}
\end{equation}

Next, for example, the first two time derivatives are%
\begin{align}
\partial _{t}n(\mathbf{r},t& \mid v)=i\left\langle \left[ H_{1}\left(
t\right) ,\widehat{n}(\mathbf{r})\right] ;0\right\rangle _{1}  \notag \\
& +\left( i\right) ^{2}\int_{0}^{t}dt^{\prime \prime }\left\langle \left[
H_{1}\left( t^{\prime \prime }\right) ,\left[ H_{1}\left( t\right) ,\widehat{%
n}(\mathbf{r})\right] \right] ;t^{\prime \prime }\right\rangle _{1}
\label{a.11}
\end{align}%
\begin{align}
\partial _{t}^{2}n(\mathbf{r},t& \mid v)=i\left\langle \left[ \partial
_{t}V_{1}\left( t\right) ,\widehat{n}(\mathbf{r})\right] ;0\right\rangle _{1}
\notag \\
& +\left( i\right) ^{2}\left\langle \left[ H_{0}\left( t\right) ,\left[
H_{1}\left( t\right) ,\widehat{n}(\mathbf{r})\right] \right] ;t\right\rangle
_{1}.  \label{a.12}
\end{align}%
The first two derivatives at $t=0$ are now readily identified. 
\begin{equation}
\partial _{t}n(\mathbf{r},t\mid v)\mid _{t=0}=i\left\langle \left[
H_{1}\left( 0\right) ,\widehat{n}(\mathbf{r})\right] ;0\right\rangle _{1}.
\label{a.14}
\end{equation}%
and%
\begin{align}
\partial _{t}^{2}n(\mathbf{r},t& \mid v)\mid _{t=0}=i\left\langle \left[
\partial _{t}V_{1}\left( t\right) \mid _{t=0},\widehat{n}(\mathbf{r})\right]
;0\right\rangle _{1}  \notag \\
& +\left( i\right) ^{2}\left\langle \left[ H_{1}\left( 0\right) ,\left[
H_{1}\left( 0\right) ,\widehat{n}(\mathbf{r})\right] \right] ;0\right\rangle
_{1}.  \label{a.15}
\end{align}%
Eq. (\ref{a.14}) determines the initial value $v_{1}(\mathbf{r}^{\prime },0)$
\begin{equation}
\int d\mathbf{r}^{\prime }v_{1}(\mathbf{r}^{\prime },0)\chi _{1}\left( 
\mathbf{r,r}^{\prime }\right) =\partial _{t}n(\mathbf{r},t\mid v)\mid
_{t=0}-i\left\langle \left[ \left( K_{1}+U_{1}\right) ,\widehat{n}(\mathbf{r}%
)\right] ;0\right\rangle _{1},  \label{a.16}
\end{equation}%
where $\chi \left( \mathbf{r,r}^{\prime }\right) $ is the static response
function 
\begin{equation}
\chi _{1}\left( \mathbf{r,r}^{\prime }\right) =i\left\langle \left[ \widehat{%
n}(\mathbf{r}^{\prime }),\widehat{n}(\mathbf{r})\right] \right\rangle _{1}.
\label{a.17}
\end{equation}%
The initial state $\rho _{1}$ is taken to be independent of $v_{1}(\mathbf{r}%
,0)$ so that Eq. (\ref{a.16}) is a linear equation for $v_{1}(\mathbf{r}%
^{\prime },0)$. In van Leeuwen's original theorem, this is interpreted as a
requirement that the average current densities of the two systems must be
the same for the initial state, using the continuity equation. Here it is
seen that this can be imposed by the choice of $v_{1}(\mathbf{r},0)$. Next,
equation Eq. (\ref{a.15}) determines the first derivative of $v_{1}(\mathbf{r%
}^{\prime },t)$%
\begin{equation}
\int d\mathbf{r}^{\prime }\partial _{t}v_{1}(\mathbf{r}^{\prime },t)\mid
_{t=0}\chi _{1}\left( \mathbf{r,r}^{\prime }\right) =\partial _{t}^{2}n(%
\mathbf{r},t\mid v)\mid _{t=0}-i\left\langle \left[ H_{1}\left( 0\right) ,i%
\left[ H_{1}\left( 0\right) ,\widehat{n}(\mathbf{r})\right] \right]
;0\right\rangle _{1}  \label{a.18}
\end{equation}%
All ingredients on the right side of this equation are known from the first
two equations, (\ref{a.10a}) and (\ref{a.16}).

The structure of Eq. (\ref{a.18}) is similar for all higher derivatives as
well. Return to Eq. (\ref{a.11}) and differentiate it $m+1$ times at $t=0$,
\ for $m>0$%
\begin{align*}
\partial_{t}^{m+1}n(\mathbf{r},t & \mid v)\mid_{t=0}=\partial_{t}^{m}Tri%
\left[ H_{1}\left( t\right) ,\rho_{1}(t)\right] \widehat {n}(\mathbf{r}%
)\mid_{t=0} \\
& =\sum_{p=0}^{m}\frac{m!}{p!\left( m-p\right) !}Tri\left[ \partial
_{t}^{m-p}V_{1}\left( t\right) ,\partial_{t}^{p}\rho_{1}(t)\right] \widehat{n%
}(\mathbf{r})\mid_{t=0}
\end{align*}%
\begin{equation}
=\left\langle \left[ \partial_{t}^{m}V_{1},\widehat{n}(\mathbf{r})\right]
;0\right\rangle _{0}\mid_{t=0}+\sum_{p=1}^{m}\frac{m!}{p!\left( m-p\right) !}%
Tri\left[ \partial_{t}^{m-p}H_{1}\left( t\right) ,\partial_{t}^{p}\rho_{1}(t)%
\right] \widehat{n}(\mathbf{r})\mid_{t=0}  \label{a.19}
\end{equation}
Rearranging gives%
\begin{equation*}
\int d\mathbf{r}^{\prime}\partial_{t}^{m}v_{1}(\mathbf{r}^{\prime},t)%
\mid_{t=0}\chi_{1}\left( \mathbf{r,r}^{\prime}\right) =\partial_{t}^{m+1}n(%
\mathbf{r},t\mid v)\mid_{t=0}
\end{equation*}%
\begin{equation}
-\sum_{p=1}^{m}\frac{m!}{p!\left( m-p\right) !}Tri\left[
\partial_{t}^{m-p}H_{1}\left( t\right) ,\partial_{t}^{p}\rho_{1}(t)\right] 
\widehat {n}(\mathbf{r})\mid_{t=0}  \label{a.20}
\end{equation}
The highest derivative of the second term on the right side is of order $m-1$
and hence denotes a quantity depending on \ known derivatives of lower order
than $m$.

The argument above constitutes a demonstration of the \textit{existence} of $%
v_{1}(\mathbf{r},t)$ in the domain of analyticity of the chosen density
about $t=0$, subject to constraints on the initial state and the
invertibility of $\chi_{1}\left( \mathbf{r,r}^{\prime}\right) $. The
argument also can be used to demonstrate \textit{uniqueness}, as follows.
Consider two systems that are the same except for their external potentials 
\begin{equation}
H\left( t\right) =K+U+V(t),\hspace{0.25in}H_{1}\left( t\right) =K+U+V_{1}(t).
\label{a.21}
\end{equation}
If it is assumed both potentials give the same density, then the
construction of their derivatives given above can be applied to each
potential. The result is that the equations for $\partial_{t}^{m}v(\mathbf{r}%
,t)\mid_{t=0}$ and for $\partial_{t}^{m}v_{1}(\mathbf{r},t)\mid_{t=0}$ are
the same (up to a constant). Consequently $v(\mathbf{r},t)$ and $v_{1}(%
\mathbf{r},t)$ are the same (they can differ by a function of time $c(t)$
since the Liouville-von Neumann equation is invariant under such a change).
In summary, there is a one-to-one relationship of the density and the
single-particle potential for a given system.

\bigskip

\section{Solution to Markov kinetic equation}

\label{appB}A formal solution to the kinetic equation, Eq. (\ref{3.17}) for $%
\phi\left( 1,\mathbf{r}^{\prime};t\right) $ is 
\begin{equation}
\phi\left( 1,\mathbf{r}^{\prime};t\right) =\phi_{KS}\left( 1,\mathbf{r}%
^{\prime};t\right) -\int_{0}^{t}dt^{\prime}e^{-iH_{KS}\left( t-t^{\prime
}\right) }\int d\mathbf{r}_{1}d\mathbf{r}_{2}\mathcal{V}_{ee}\left( \mathbf{r%
}_{1},\mathbf{r}_{2}\right) I(1,\mathbf{r}_{1},\mathbf{r}_{2},\mathbf{r}%
^{\prime},t^{\prime})e^{iH_{KS}\left( t-t^{\prime}\right) },  \label{c.1}
\end{equation}
where 
\begin{equation}
\phi_{KS}\left( 1,\mathbf{r}^{\prime};t\right) =e^{-iH_{KS}t}\phi\left( 1,%
\mathbf{r}^{\prime};0\right) e^{iH_{KS}t},  \label{c.2}
\end{equation}%
\begin{equation}
H_{KS}=\sum_{i=1}^{N_{e}}h_{KS}(i),\hspace{0.25in}h_{KS}(i)=\frac{p_{i}^{2}}{%
2m}+v_{KS}\left( \mathbf{q}_{i},\left\{ \mathbf{R}\right\} \right) .
\label{c.2a}
\end{equation}
Recall that $v_{KS}\left( \mathbf{q}_{i},\left\{ \mathbf{R}\right\} \right) $
is a functional of the initial equilibrium density and therefore $H_{KS}$ is
time independent. Also, 
\begin{align}
I(1,\mathbf{r}_{1},\mathbf{r}_{2},\mathbf{r}^{\prime},t^{\prime}) & \equiv
Tr_{2}i[\delta\left( \mathbf{r}_{1}-\mathbf{q}_{1}\right) \delta\left( 
\mathbf{r}_{2}-\mathbf{q}_{2}\right) ,f^{(1)}(1)\phi\left( 2,\mathbf{r}%
^{\prime};t^{\prime}\right) ]  \notag \\
& =i[\delta\left( \mathbf{r}_{1}-\mathbf{q}_{1}\right)
,f^{(1)}(1)]Tr_{2}\phi\left( 2,\mathbf{r}^{\prime};t^{\prime}\right)
\delta\left( \mathbf{r}_{2}-\mathbf{q}_{2}\right)  \notag \\
& =-\phi\left( 1,\mathbf{r}_{1};0\right) \chi\left( \mathbf{r}_{2},\mathbf{r}%
^{\prime};t^{\prime}\right) .  \label{c.3}
\end{align}
The definition of $\phi\left( 1,\mathbf{r};0\right) $ in Eq. (\ref{3.11})
and of $\chi\left( \mathbf{r},\mathbf{r}_{2};t^{\prime}\right) $ in Eq. (\ref%
{3.10}) has been used in the last line.

The response function is given by Eq. (\ref{3.7}) 
\begin{equation}
\chi\left( \mathbf{r},\mathbf{r}^{\prime};t\right) =Tr_{1}\delta\left( 
\mathbf{r}-\mathbf{q}_{1}\right) \phi\left( 1,\mathbf{r}^{\prime};t\right) .
\label{c.3a}
\end{equation}
With Eq. (\ref{c.1}) this becomes 
\begin{align}
\chi\left( \mathbf{r},\mathbf{r}^{\prime};t\right) & =\chi_{KG}\left( 
\mathbf{r},\mathbf{r}^{\prime};t\right) +\int_{0}^{t}dt^{\prime}\int d%
\mathbf{r}_{1}d\mathbf{r}_{2}\mathcal{V}_{ee}\left( \mathbf{r}_{1},\mathbf{r}%
_{2}\right) Tr_{1}\delta\left( \mathbf{r}-\mathbf{q}_{1}\right)
e^{-iH_{KS}\left( t-t^{\prime}\right) }\phi\left( 1,\mathbf{r}_{1};0\right)
e^{iH_{KS}\left( t-t^{\prime}\right) }\chi\left( \mathbf{r}_{2},\mathbf{r}%
^{\prime};t^{\prime}\right)  \notag \\
& =\chi_{KG}\left( \mathbf{r},\mathbf{r}^{\prime};t\right)
+\int_{0}^{t}dt^{\prime}\int d\mathbf{r}_{1}d\mathbf{r}_{2}\chi_{KG}\left( 
\mathbf{r},\mathbf{r}_{1};t-t^{\prime}\right) \mathcal{V}_{ee}\left( \mathbf{%
r}_{1},\mathbf{r}_{2}\right) \chi\left( \mathbf{r}_{2},\mathbf{r}%
^{\prime};t^{\prime}\right)  \label{c.4}
\end{align}
where the Kubo-Greenwood response function is%
\begin{equation}
\chi_{KG}\left( \mathbf{r},\mathbf{r}^{\prime};t\right) =Tr_{1}\delta\left( 
\mathbf{r}-\mathbf{q}_{1}\right) \phi_{KS}\left( 1,\mathbf{r}^{\prime
};t\right) .  \label{c.5}
\end{equation}

\end{document}